\documentclass[showpacs,aps,prl,graphicx,twocolumn]{revtex4}%
\usepackage{graphicx}

\begin{document}

\title{Multiparty Quantum Secret Report}

\author{ Fu-Guo Deng$^{1,2,3}$\footnote{Email: fgdeng@bnu.edu.cn },
 Xi-Han Li$^{1,2}$,
  Chun-Yan Li$^{1,2}$, Ping Zhou$^{1,2}$,
Yu-Jie Liang$^{1,2}$ and  Hong-Yu Zhou$^{1,2,3}$ }

\address{$^1$ The Key Laboratory of Beam Technology and Material
Modification of Ministry of Education, Beijing Normal University,
Beijing 100875,
People's Republic of China\\
$^2$ Institute of Low Energy Nuclear Physics, and Department of
Material Science and Engineering, Beijing Normal University,
Beijing 100875, People's Republic of China\\
$^3$ Beijing Radiation Center, Beijing 100875,  People's Republic of
China}
\date{\today }


\begin{abstract}

A multiparty quantum secret report scheme is proposed with quantum
encryption. The boss Alice and her $M$ agents first share a
sequence of ($M$+1)-particle Greenberger--Horne--Zeilinger (GHZ)
states that only Alice knows which state each ($M$+1)-particle
quantum system is in. Each agent exploits a controlled-not (CNot)
gate to encrypt the travelling particle by using the particle in
the GHZ state as the control qubit. The boss Alice decrypts the
travelling particle with a CNot gate after performing a $\sigma_x$
operation on her particle in the GHZ state or not. After the GHZ
states (the quantum key) are used up, the parties check whether
there is a vicious eavesdropper, say Eve, monitoring the quantum
line, by picking out some samples from the GHZ states shared and
measure them with two measuring bases. After confirming the
security of the quantum key, they use the GHZ states remained
repeatedly for next round of quantum communication. This scheme
has the advantage of high intrinsic efficiency for qubits
and the total efficiency.

\end{abstract}
\pacs{03.67.Dd, 03.67.Hk, 03.65.Ud} \maketitle

The main task of quantum communication is to transmit message
securely between the authorized parties and its implement can be
completed by the transmission of quantum states.\cite{book,gisin}
The features of quantum mechanics such as the uncertainty principle,
quantum correlations and non-locality play an important role for the
security of message.\cite{book} Any of eavesdropper's action will
inevitably leave a trace in the result and the parties of
communication will detect the eavesdropping by comparing the results
of the samples chosen randomly and then abandon all the results
transmitted to avoid leaking the secret message.\cite{gisin} The
security of transmission for quantum communication is embodied to
the fact that the sender Alice and receiver Bob can find out the
malicious or dishonest action of others. Thus quantum communication
is secure for generating key
\cite{gisin,bb84,ekert91,bbm92,LongLiu,CORE} and secret message
splitting
\cite{HBB99,KKI,Guo,longQSS,yanpra,Peng,dengQSTS,dengcontrolled,Zhang}
among the parties of communication. Experimental implement for
quantum key distribution has been studied by a lot of groups
\cite{gisin}

Recently, the concept of quantum secure direction communication
(QSDC), which is used not for distributing key but just for
transmitting secret message directly, was proposed and actively
pursued.
\cite{imo1,imo2,beige,bf,two-step,QOTP,Wangc,zhangzj,yan,Gao,zhangsPRA,cai,caiPRA,song,linetwork}
Shimizu and Imoto \cite{imo1,imo2} and Beige {\it et al.}
\cite{beige} proposed some QSDC schemes, in which the secret message
can be read out after at least one bit of additional classical
information is transmitted for each qubit. In 2002, Bostr\"{o}m and
Felbinger \cite{bf} introduced an insecure ping-pong scheme for
direct communication and key distribution \cite{attack1}.
Subsequently, we \cite{two-step} put forward a two-step QSDC
protocol with Einstein--Podolsky--Rosen (EPR) pairs transmitted in
quantum data block and another one based on a sequence of polarized
single photons.\cite{QOTP} Wang {\it et al.}\cite{Wangc} introduced
a QSDC protocol with high-dimension quantum superdense coding.

In 2005, Gao {\it et al.}\cite{yannetwork} proposed a scheme for
secure direct communication between the central party and the
other $M$ parties with Greenberger--Horne--Zeilinger (GHZ) states.
In their scheme, the $M$ parties can communicate the central party
securely with entanglement swapping. That is, they can send their
secret message directly to the central party. We call this model
the multiparty quantum secret report (MQSR) as it is used to complete
the task that the agents report their secret message to the Boss
in one-way direction. In 2006, Jin {\it et al.} \cite{JinPLA}
introduced another MQSR scheme with GHZ states following some
ideas in quantum dense coding.\cite{BW}

In this Letter, we present an MQSR scheme with GHZ states via
quantum encryption. The boss Alice and her $M$ agents first share a
sequence of ($M$+1)-particle quantum systems  with a decoy
technique. The states of the quantum systems are kept by Alice and
any other one does not know. Each agent exploits a controlled-not
(CNot) gate to encrypt the travelling particle $T$ by using the
particle in the GHZ state as the control qubit. The boss Alice
decrypts the travelling particle $T$ with a CNot gate after
performing a $\sigma_x$ operation on her particle in the GHZ state
or not, according to the fact that her particle is correlated or
anti-correlated to that controlled by the agent. After the GHZ
states are used up, the parties check whether there is a vicious
eavesdropper monitoring the quantum line by selecting some samples
and measuring them with two measuring bases (MBs), X-MB and Y-MB.
After confirming the security of the quantum key, they use the GHZ
states remained repeatedly for next round of quantum communication.
 With quantum storage technique, this scheme has
high intrinsic efficiency for qubits and the total efficiency.

Now, let us describe the principle of our MQSR scheme in detail
with $M$ agents, say Bob$_r$ ($r=1,2,\cdots, M$).

All the ($M$+1)-particle GHZ states can be described by
\cite{dengQSTS}
\begin{eqnarray}
\vert G_{\underbrace{j\cdots k}_M+}\rangle &=&\frac{1}{\sqrt{2}}
(\vert 0 \underbrace{j\cdots k}_M\rangle + \vert 1
\underbrace{\bar{j}\cdots \bar{k}}_M\rangle)_{AB_1B_2 \cdots B_M},\nonumber\\
\vert G_{\underbrace{j\cdots k}_M-}\rangle &=&\frac{1}{\sqrt{2}}
(\vert 0 \underbrace{j\cdots k}_M\rangle - \vert 1
\underbrace{\bar{j}\cdots \bar{k}}_M\rangle)_{AB_1B_2
\cdots B_M},\nonumber\\
\end{eqnarray}
where $j,k\in \{0, 1\}$, $\bar{j}$ and $\bar{k}$ are the
counterparts of the binary numbers $j$ and $k$, respectively; $\vert
0\rangle$ and $\vert 1\rangle$ are the two eigenstates of the Z-MB.

Our MQSR scheme can work with the  following steps:

(S1) The Boss Alice shares a sequence of GHZ-state quantum systems,
say $S_q$, with all her agents Bob$_r$ ($r=1,2,\cdots, M$) privately
and securely. That is, Alice prepares each quantum system randomly
in one of the GHZ states $\{\vert G_{j\cdots k+}\rangle\}$ and keeps
the secret information about the states for any one. She sends the
particles $B_r$ to the agent Bob$_r$ and always controls the
particle $A$ for each quantum system. The parties can analyse the
security of the GHZ states by the following method. (a) Alice picks
out a sufficiently subset of the states as the samples, say $s_{e1}$
for eavesdropping check, similar to Ref. \cite{two-step}. (b) The
agents choose X-MB $=\{\vert +x\rangle=\frac{1}{\sqrt{2}}(\vert
0\rangle + \vert 1\rangle),\vert -x\rangle=\frac{1}{\sqrt{2}}(\vert
0\rangle - \vert 1\rangle)\}$ or Y-MB $=\{\vert
+y\rangle=\frac{1}{\sqrt{2}}(\vert 0\rangle + i\vert 1\rangle),\vert
-y\rangle=\frac{1}{\sqrt{2}}(\vert 0\rangle - i\vert 1\rangle)\}$
randomly for their particles in the samples $s_{e1}$. They tell
Alice their MBs for their measurements, and then Alice measures her
particles with the correlated MBs. For example, if the state of the
GHZ-state quantum system is $\vert G_{\underbrace{0\cdots
0}_M+}\rangle =\frac{1}{\sqrt{2}} (\vert 0 \underbrace{0\cdots
0}_M\rangle + \vert 1 \underbrace{1\cdots 1}_M\rangle)_{AB_1B_2
\cdots B_M}$, Alice chooses X-MB for her particle if the number of
the agents who choose Y-MB for their measurements on their particles
is even, otherwise Alice chooses Y-MB for her measurement. The other
cases is the same as this one with or without a little of
modification. (c) The agents publish their results of their
measurements, and Alice analyses the error rate of the samples. If
the error rate is very low, they can conclude that the quantum key,
the $N$ ordered GHZ states is created securely. Otherwise, they will
discard the results and repeat the quantum communication from the
beginning.

For preventing some dishonest agents from stealing the information
about the GHZ states with a fake signal, Alice adds some decoy
photons in the sequence $S_q$. That is, Alice prepares some
photons randomly in the states $\{\vert +x\rangle, \vert
-x\rangle, \vert +y\rangle, \vert -y\rangle\}$, with which she
replaces some particles $B_r$ in $S_q$. In this way, the dishonest
agent cannot eavesdrop the information for the quantum key $S_q$
with intercepting-resending attack and cheat. Let us use a simple
case as an example to describe the principle of this
eavesdropping. Assume that there are only two agents, say Bob$_1$
and Bob$_2$, and the three-particle quantum system is in the state
$\vert \Psi\rangle_0\equiv\vert
GHZ\rangle_{00+}=\frac{1}{\sqrt{2}}(\vert 000\rangle + \vert
111\rangle)_{AB_1B_2}$. If Bob$_1$ is a dishonest man, he can
intercept the particle $B_2$ when it runs from Alice to Bob$_2$.
Instead, Bob$_1$ prepares a Bell state $\vert
\psi^-\rangle=\frac{1}{\sqrt{2}}(\vert 01\rangle-\vert
10\rangle)_{b_1b_2}$ and sends Bob$_2$ the particle $b_2$ and
keeps the particle $b_1$. When this quantum system is chosen by
Alice as a sample for eavesdropping check, Bob$_1$ performs a
Bell-basis measurement on the particles $B_2$ and $b_1$. Then the
difference of the state $\vert \Psi\rangle_0$ and that of the
particles $AB_1b_2$ is just a local unitary operation, $I$,
$\sigma_z$, $\sigma_x$ or $i\sigma_y$,
\begin{eqnarray}
\vert \Psi\rangle_0&=& \frac{1}{2}[(\vert +x\rangle\vert +x\rangle
+ \vert
-x\rangle\vert -x\rangle)_{B_1B_2}\vert +x\rangle_A \nonumber\\
&& +(\vert +x\rangle\vert -x\rangle + \vert -x\rangle\vert
+x\rangle)_{B_1B_2}\vert -x\rangle_A] \nonumber\\
&=& \frac{1}{2}[(\vert +y\rangle\vert -y\rangle + \vert
-y\rangle\vert +y\rangle)_{B_1B_2}\vert +x\rangle_A \nonumber\\
&& +(\vert +y\rangle\vert +y\rangle + \vert -y\rangle\vert
-y\rangle)_{B_1B_2}\vert -x\rangle_A].\label{GHZ1}
\end{eqnarray}
\begin{eqnarray}
\vert \Psi\rangle_0 \otimes \vert \psi^-\rangle &=&
\frac{1}{\sqrt{2}}(\vert 000\rangle + \vert
111\rangle)_{AB_1B_2}\nonumber\\
&& \otimes\frac{1}{\sqrt{2}}(\vert
01\rangle-\vert 10\rangle)_{b_1b_2}\nonumber\\
&=& \frac{1}{2\sqrt{2}}[(\vert 001\rangle - \vert
110\rangle)_{AB_1b_2}\vert \phi^+\rangle_{B_2b_1} \nonumber\\
&& +(\vert 001\rangle + \vert
110\rangle)_{AB_1b_2}\vert \phi^-\rangle_{B_2b_1} \nonumber\\
&&- (\vert 000\rangle - \vert
111\rangle)_{AB_1b_2}\vert \psi^+\rangle_{B_2b_1} \nonumber\\
&& -(\vert 000\rangle + \vert 111\rangle)_{AB_1b_2}\vert
\psi^-\rangle_{B_2b_1}],\label{GHZswapping}
\end{eqnarray}
where
\begin{eqnarray}
\vert \phi^{\pm}\rangle = \frac{1}{\sqrt{2}}(\vert 00\rangle \pm
\vert 11\rangle), \;\;\; \vert \psi^{\pm}\rangle =
\frac{1}{\sqrt{2}}(\vert 01\rangle \pm \vert
10\rangle).\label{GHZswapping}
\end{eqnarray}
Bob$_1$ can hide the difference with cheat. Assume that the final
state of the particles $AB_1b_2$ is
\begin{eqnarray}
\vert \Psi\rangle_1 &\equiv& \vert
GHZ\rangle_{00-}=\frac{1}{\sqrt{2}}(\vert 000\rangle - \vert
111\rangle)\nonumber\\
&=& \frac{1}{2}[(\vert +x\rangle\vert -x\rangle + \vert
-x\rangle\vert +x\rangle)_{B_1b_2}\vert +x\rangle_{A} \nonumber\\
&& +(\vert +x\rangle\vert +x\rangle + \vert -x\rangle\vert
-x\rangle)_{B_1b_2}\vert -x\rangle)_{A}] \nonumber\\
&=& \frac{1}{2}[(\vert +y\rangle\vert +y\rangle + \vert
-y\rangle\vert -y\rangle)_{B_1b_2}\vert +x\rangle_A \nonumber\\
&& +(\vert +y\rangle\vert -y\rangle + \vert -y\rangle\vert
+y\rangle)_{B_1b_2}\vert -x\rangle_A].\label{fakeGHZ}
\end{eqnarray}
Bob$_1$ can publish a counterpart of the result of his measurement
for his attack when the quantum system $AB_1B_2$ is chosen for
eavesdropping check, i.e., if he measures his particle $B_1$ and
obtains his result $\vert +x\rangle$, he tells Alice that his result
is $\vert -x\rangle$. Bob$_1$'s cheat cannot be detected. The cases
with the final states $\frac{1}{\sqrt{2}}(\vert 001\rangle \pm \vert
110\rangle_{AB_1b_2}$ are the same as this one with or without a
little of modification. Fortunately, this attack does not work for
the decoy photons as its security is the same as the
Bennett-Brassard quantum key distribution protocol.
\cite{bb84,BB84security} Thus the parties can share a sequence of
GHZ states securely.

(S2) The agents use the GHZ states as the quantum key and send
their secret message to the boss Alice with a controlled-not (CNot)
gate, similar to Ref. \cite{qcryption1} (the difference is just
that the quantum key is unknown for the agents in this scheme). In
detail, a agent, say Bob$_r$, can report his secret message to the
boss Alice by means that he prepares a travelling particle $T$ in
the state $\vert \alpha\rangle_T$ $\in$ $\{\vert 0\rangle,\;\vert
1\rangle\}$, which represents the classical bit 0 and 1 and takes a
CNot operation on the particle $T$ by using the qubit $B_r$ as the
control qubit. For instance, we assume that the GHZ state in the
quantum key is
\begin{eqnarray}
\vert G\rangle =\frac{1}{\sqrt{2}} (\vert 0
j\cdots i \cdots k\rangle 
+ \vert 1 \bar{j}\cdots  \bar{i} \cdots \bar{k}\rangle)_{AB_1B_2
\cdots B_r \cdots B_M}.
\end{eqnarray}
After the encryption carried out by Bob$_r$ with a CNot operation, the
state $\vert\Psi\rangle_{s1}$ of the whole quantum system composed
of the GHZ-state particles and the travelling particle $T$ is
\begin{eqnarray}
\vert \Psi\rangle_{s1}&=& \frac{1}{\sqrt{2}} (\vert 0
j\cdots i \cdots k\rangle \vert i\oplus \alpha\rangle \nonumber\\
&+& \vert 1 \bar{j}\cdots  \bar{i} \cdots \bar{k}\rangle \vert
\bar{i}\oplus \alpha\rangle)_{AB_1B_2 \cdots B_r \cdots B_M T}.
\end{eqnarray}
Bob$_r$ sends the particle $T$ to Alice who can decrypt the qubit
$T$ with a CNot operaton on the particles $T$ and $A$  by using
her qubit $A$ as the control qubit after flipping its bit value or
not (if the qubits $B_r$ and $A$ is anti-correlated, Alice should
first operates the qubit $A$ with the Pauli operation $\sigma_x$).
The agent Bob$_r$ continues the quantum communication until the
quantum key is used up.

In order to estimate the error rate of the transmission for the
$T$ sequence which is composed of all the travelling particles
sent by Bob$_r$, he should insert some sample particles $s_{e3}$
in the $T$ sequence before it runs in the quantum line. The number
of the particles in $s_{e3}$ is not required to be very large, but
enough for the statistical analysis. As any eavesdropper does not
know the information about the GHZ states, he cannot decrypt the
encryption done by Bob$_r$. Thus the security of the quantum
communication between Alice and Bob$_r$ is the same as that in
quantum one-time pad.\cite{gisin}

(S3) The agents and Alice use repeatedly the quantum key
conditionally, similar to Ref. \cite{qcryption2}. That is, they
first check the security of the quantum key by choosing a subset of
the GHZ states and measuring them with the two MBs, X-MB and Y-MB,
same as that in the process for creating the quantum key. If Alice
confirms that the quantum key is secure, the parties use the GHZ
states remained again. As the amount of the samples chosen for
checking eavesdropping is negligible comparing with  all the GHZ
states, the quantum key is used repeatedly in the next round if
there is no one monitoring the quantum line.

For saving the GHZ states in the quantum key in the process for
eavesdropping check, Alice and Bob$_r$ can first choose the GHZ
states used for encrypting the sample particles $s_{e3}$ as the
samples for checking the security of the quantum key. Then they
choose some other GHZ states in the sequence $S_q$ for the
eavesdropping check.

(S4) The agents and Alice repeat the steps (2) and (3) until their
task is accomplished.

Different from the classical key in one-time pad crypto-system,
the quantum key in this MQSR scheme can be repeatedly used except
for the states chosen for checking eavesdropping. For the
eavesdropper Eve the quantum key is randomly in one of the two
eigenstates of her measuring operation, and then she cannot
decrypt the qubits in the quantum key. Moreover, her action will
destroy the correlation of the particles in a GHZ state and will
be detected by the authorized parties. The randomness of the
quantum key for the eavesdropper ensures the security of the
secret message transmitted.

With quantum storage technique,\cite{storage} this MQSR scheme has
the advantage of high intrinsic efficiency for qubits $\eta_q\equiv
\frac{q_u}{q_t}$ since almost all qubits are useful in principle for
transmitting the secret message. Here $q_u$ and $q_t$ are the qubits
useful and total qubits for the transmission, respectively. The
total efficiency $\eta_t$ in this scheme approaches 100\% as the
classical information exchanged is unnecessary except for the
eavesdropping check. Here $\eta$ is defined as \cite{Cabello}
\begin{eqnarray}
\eta_t=\frac{q_u}{q_t + b_t},
\end{eqnarray}
where $b_t$ is the classical bits exchanged between the parties in
the quantum communication.

In summary, we have presented a scheme for multiparty quantum secret
report with quantum encryption. In this scheme, the Boss Alice first
shares a sequence of GHZ states as the quantum key, and then the
agents use a CNot operation to encrypt their message and then send
it to the Boss. The states of the GHZ quantum systems are unknown
for any of the agents, and the quantum communication is secure same
as a quantum one-time pad crypto-system. They can use the quantum
key repeatedly except for the GHZ states used for check
eavesdropping if there is no-one monitoring the quantum line. This
scheme has advantage of having high intrinsic efficiency for qubits
$\eta_q$ and the total efficiency $\eta_t$ as almost all the qubits
are useful for carrying the secret message and the classical
information exchanged is unnecessary for the qubits but those chosen
for check eavesdropping. On the other hand, the parties should have
the capability of storing the GHZ states.

This work is supported by the National Natural Science Foundation of
China under Grant No 10447106, and Beijing Education Committee under
Grant No XK100270454.

\end{document}